\documentclass[fleqn,twoside]{article}
\usepackage{espcrc2}

\usepackage{graphicx}

\newcommand{\AmS}{{\protect\the\textfont2
  A\kern-.1667em\lower.5ex\hbox{M}\kern-.125emS}}

\title{PC Clusters for Lattice QCD}

\author{D.J. Holmgren\address[FNAL]{Computing Division, Fermi National Accelerator Laboratory,
        P.O. Box 500, Batavia, IL, USA}%
}
       
\begin{document}

\begin{abstract}
In the last several years, tightly coupled PC clusters have become widely
applied, cost effective resources for lattice gauge computations.  This paper
discusses the practice of building such clusters, in particular balanced
design requirements.  I review and quantify the improvements over time of key
performance parameters and overall price to performance ratio.  Applying these
trends and technology forecasts given by computer equipment manufacturers, I
predict the range of price to performance for lattice codes expected in the
next several years.

\vspace{1pc}
\end{abstract}

\maketitle

\section{INTRODUCTION}
The simulation codes of lattice gauge theory require substantial computing
resources in order to calculate various matrix elements with sufficient
precision to test the Standard Model against emerging experimental
measurements.  Historically, these codes have demanded the use of large
supercomputers at significant cost.  Both general purpose commercial
supercomputers and custom, or ``purpose-built'', supercomputers have been
employed.  Traditional supercomputers came with very high prices.  The price
of purpose-built supercomputer hardware was lower, but the design and
construction of such machines required significant amounts of engineering and
physicist manpower.

In the last half decade, the performance of commodity computing equipment has
increased to the point that tightly coupled clusters of such machines can
compete with traditional supercomputers in capacity (lattice size) and
throughput (MFlop/sec), and with purpose-built supercomputers in
price/performance.  Commodity systems have been so successful across a wide
spectrum of applications in many academic fields, that more than half of the
supercomputers listed on the ``Top500''~\cite{top500} supercomputer list are
clusters.

In this paper, I discuss the requirements placed on clusters by lattice QCD
codes and the historical performance trends of commodity computing equipment
for meeting those requirements.  Extrapolating from these trends, together
with vendor roadmaps, allows prediction of the performance and
price/performance of reasonable cluster designs in the next few years. 

\section{DESIGNING BALANCED SYSTEMS}
Inversion of the Dirac operator ({\em Dslash}) is the most computationally
intensive task of lattice codes.  The improved staggered action ({\em asqtad})
will be used throughout this paper for quantitative examples.  During each
iteration of the inversion of the improved staggered {\em Dslash}, eight sets
of SU(3) matrix-vector multiplies occur using nearest and next-next-nearest
neighbor spinors.  When domain decomposition is used on a cluster, ideally
these floating point operations overlap with the communication of the
hyper-surfaces of the sub-lattices held on neighboring nodes.  Using global
sums, the results of these sweeps over the full lattice are accumulated and
communicated to all nodes in order to modify the spinors for the next
iteration.

{\em Dslash} inversion throughput depends upon the floating point performance
of the processors, the bandwidth available for reading operands from memory,
the throughput of the I/O bus of the cluster nodes, and the bandwidth and
latency of the network fabric connecting the computers.  On any cluster, one
of these factors will be the limiting factor which dictates performance for a
given problem size.  Minimization of price/performance requires designs which
balance these factors.

\subsection{Floating Point Performance}
Most floating point operations in lattice codes occur during SU(3)
matrix-vector multiplies.  For operands in cache, the throughput of these
multiplies is dictated by processor clock speed and the capabilities of the
floating point unit.  Table~\ref{matvec} shows the performance of
matrix-vector kernels on four Intel processors introduced since the year 2000.
The ``C'' language kernels used are from the MILC~\cite{MILC} code.  The use
of SIMD instructions on Intel-brand and compatible CPUs, as suggested by Csikor
{\em et al.}~\cite{Fodor} for AMD K6-2 CPUs and implemented for the
Intel SSE unit by L\"{u}scher~\cite{Luscher}, can give significant performance
improvements.  Table~\ref{matvec} lists the performance of two styles of SSE
implementation.  The first, site wise, uses a conventional data layout scheme
with the real and imaginary pieces of individual matrix and vector elements
adjacent in memory.  The second, fully vectorized, follows
Pochinsky's~\cite{Pochinsky} practice of placing the real components of the
operands belonging to four consecutive lattice sites consecutively in memory,
followed by the four imaginary components.  Whereas site wise implementations
require considerable shuffling of operands in the SSE registers in order to
perform complex multiplies, the fully vectorized form requires only loads,
stores, multiplies, additions, and subtractions.

\begin{table}[t]
\caption{SU(3) matrix-vector multiply performance.  Results are given in MFlop/sec.}
\label{matvec}
\begin{tabular}{l c c c}
\hline
Processor & ``C'' & Site-Wise & Vector \\
\hline
1.5 GHz Xeon & 860 & 1710 & 5450 \\
2.4 GHz Xeon & 1310 & 2760 & 8190 \\
2.8 GHz P4 & 1530 & 3220 & 9560 \\
2.8 GHz P4E & 1210 & 2710 & 7400 \\
\hline
\end{tabular}
\vspace{-0.5cm}
\end{table}

\subsection{Memory Performance}
The bandwidth of access to main memory by processors depends upon the width
and the clock speed of the data bus.  Intel and compatible {\tt ia32} architecture
processors use 64-bit data buses exclusively.  The effective speed of the
so-called {\em front side bus}, or FSB, has increased from 66 MHz in the
mid-90's, to 800 MHz today.  The corresponding peak memory bandwidths have
increased from 528 MB/sec to 6400 MB/sec.  According to Intel roadmaps,
processors with 1066 MHz FSB and 8530 MB/sec peak bandwidths will be available
by November of 2004.  The doubling time for the exponential fit to these
bandwidths is 1.87 years.  The doubling for achievable bandwidth, measured
using the STREAMS~\cite{streams} benchmark, is 1.71 years.  With SSE
optimizations, the achieved doubling time decreases to 1.49 years.

From memory bandwidth measurements, using tools such as STREAMS, an estimate
of the throughput of SU(3) matrix-vector multiply kernels can be made in the
case in which all operands come from main memory, typical for lattice QCD
codes.  For single precision calculations, each matrix-vector multiply
requires 96 input bytes, 24 output bytes, and 66 floating point operations.
The throughput is given by this flop count divided by the memory access speed,
weighted appropriately according to read and write rates.  Table~\ref{membw}
shows the main memory matrix-vector throughput for six generations of {\tt ia32}
processor, along with the conventional and SSE assisted read and write rates.
Comparing Table~\ref{membw} to Table~\ref{matvec} clearly shows that memory
bandwidth constrains lattice QCD code performance.

\begin{table}[t]
\caption{Memory bandwidth, and SU(3) matrix-vector throughput. 
FSB is given in MHz.  Read and write rates are in MB/sec, measured using
SSE-assisted code except for the PPro. The final column gives inferred SU(3)
matrix-vector throughput in MFlop/sec.}
\label{membw}
\begin{tabular}{l c c c c}
\hline
Processor & FSB & Read & Write & M-V \\
\hline
PPro 200 MHz & 66 & 98 & 98 & 54 \\
P-III 733 MHz & 133 & 880 & 1000 & 500 \\
P4 1.4 GHz & 400 & 2070 & 2120 & 1140 \\
Xeon 2.4 GHz & 400 & 2260 & 1240 & 1070 \\
P4 2.8 GHz & 800 & 4100 & 3990 & 2240 \\
P4E 2.8 GHz & 800 & 4560 & 2810 & 2230 \\
\hline
\end{tabular}
\vspace{-0.5cm}
\end{table}

\subsection{Communications Requirements and I/O Bus Performance}
Gottlieb~\cite{gottlieb} has presented a very useful model for understanding
the communications requirements of lattice QCD code.  Modified for this paper
for the improved staggered action, this model assumes a hypercubic lattice
evenly divided among the nodes of a cluster, with inter-node communications
occurring along all 4 directions.  The size of the sub-lattice stored on each
node, along with the requirement of the algorithm that data from the three
outermost hyperplanes in each direction be communicated between neighboring
nodes, gives the size of the messages interchanged during each iteration
of the {\em Dslash} inverter.  Therefore, for any assumed {\em Dslash}
performance and sub-lattice size, one can easily determined the necessary
communications bandwidth.  The required bandwidth increases with decreasing
sub-lattice size, and increases with increasing {\em Dslash} throughput.

\begin{figure}[t]
\includegraphics[width=18pc,height=12pc]{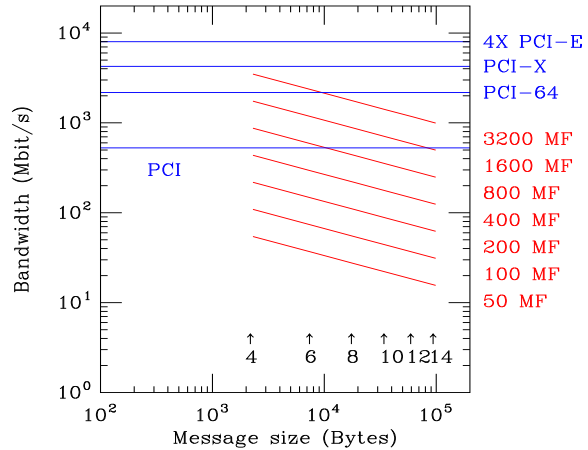}
\caption{
The diagonal lines show the required communication bandwidths as a function
of message size and assumed {\em Dslash} performance (MFlop/sec).  The
marked message sizes, $L$, correspond to sub-lattices of size $L^4$.  The
horizontal lines show burst rates for these buses:  PCI 32-bit, 33 MHz (132
MB/sec), PCI 64-bit, 66 MHz (528 MB/sec), PCI-X 64-bit, 133 MHz (1064 MB/sec),
and 4X PCI-E (2000 MB/sec).
\vspace{-0.50cm}
}
\label{gottlieb}
\end{figure}

The maximum communications rate between nodes in a cluster is limited by the
smaller of the I/O bus and network bandwidths.  Figure~\ref{gottlieb} shows
the required bandwidths from the model as a function of message size for a
variety of assumed {\em Dslash} throughputs.  The labeled horizontal lines
show the burst communications rates of various I/O buses, from the 132 MB/sec
rate for the 32-bit, 33 MHz PCI bus of the mid-90's, to the 2000 MB/sec rate
for the four-lane PCI Express (PCI-E) introduced in 2004.  For any of the I/O
architectures shown, the achievable communications rate will be no more than
perhaps 75\% of these burst rates.  This plot shows that for current
processors, capable of achieving 800 to 1600 MFlop {\em Dslash} throughput,
PCI-X (64-bit, 133 MHz) buses are sufficient.  Furthermore, currently
available sixteen lane PCI-E will be more than sufficient for at least six
more years, when processors could achieve at least 10 GFlop throughput.


\subsection{Network Fabric Performance}
A number of network fabrics exist with sufficient performance for lattice QCD
clusters.  These include gigabit ethernet employing switches or toroidal
meshes~\cite{fodor2,watson}, Myrinet, Quadrics, SCI, and Infiniband.  Gigabit
ethernet meshes of high dimensionality have the advantage of very low cost,
but the disadvantages of large numbers of cables, the need for custom
software, and sensitivity to node failures.  SCI, another multi-dimensional
toroidal mesh, is robust against node failures but at higher cost than gigabit
ethernet.  Myrinet, Quadrics, and Infiniband all employ switched fabrics and
have been used in large (order 1000) node clusters in fields outside of
lattice QCD.

\begin{figure}[t]
\includegraphics[width=18.5pc,height=12pc]{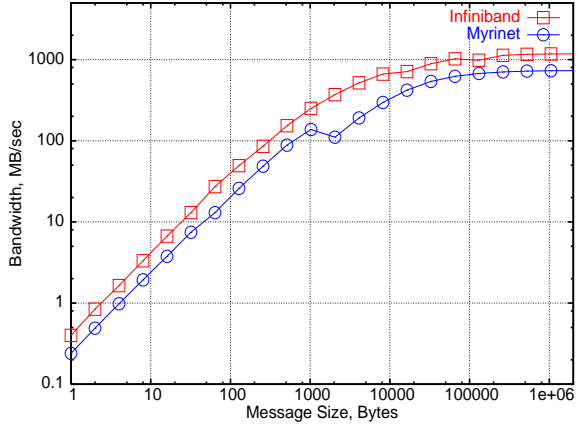}
\caption{Network bandwidth as a function of message size, measured using the MPI
Netpipe benchmark.
\vspace{-0.50cm}
}
\label{ib_myr}
\end{figure}

Examples of communications performance for Myrinet (LANai9, PCI64B) and
Infiniband (PCI-X) networks are shown in Fig.~\ref{ib_myr}.  Typical for all
fabrics is the bandwidth saturation at large message sizes, limited by either
the I/O bus or the network itself, and the steady decrease in bandwidth with
decreasing message size because of the delay (latency) necessary to setup and
process a communication.  {\em Dslash} inversion usually involves message
sizes of order 1000 bytes or higher.  The dispersion of bandwidth with message
size determines how small a sub-lattice may be employed.  For a fixed problem
size, increasing the number of nodes decreases the time required to perform
the calculation when the parallel computer is limited by floating point
performance or memory bandwidth.  However, since bandwidth also declines with
the smaller message sizes, as the number of nodes increases eventually the
network will become the limiting performance factor.

\begin{figure}[h]
\includegraphics[width=18pc,height=12pc]{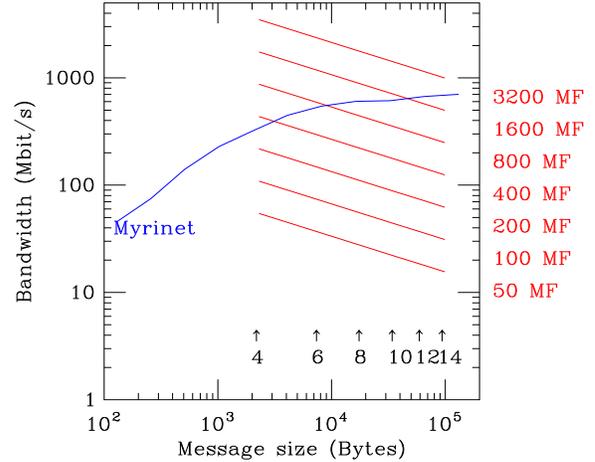} 
\caption{The Myrinet dispersion curve of Fig.~\ref{ib_myr} is superimposed on the
communications model of Fig.~\ref{gottlieb} in order to estimate network
limited performance as a function of processor performance and sub-lattice
size.}
\label{gott_myr}
\end{figure}

A rough estimate of this cutoff for a given sub-lattice size may be obtained
by superimposing the network bandwidth dispersion curve onto the model curves
of the last section.  See Fig.~\ref{gott_myr} for an example using Myrinet,
where the dispersion curve was obtained using a two-node MPI
Netpipe~\cite{netpipe} benchmark.  For this network, message sizes of at least
$10^4$ bytes are required for 800 MFlop {\em Dslash} throughput.  Note,
however, that this cutoff estimate is an optimistic upper bound.  Unlike
Netpipe, there is contention for both the I/O and memory buses when lattice
QCD code runs.  I/O bus contention results from in-bound and out-bound
messages occurring simultaneously.  Competition for the memory bus results
from the overlap of communications with computation.

As with floating point performance, memory bandwidth, and I/O bus capability,
network fabrics have steadily improved in performance over the last decade.
Ethernet speeds have increased from 10 Mbit/sec in the early '90s to 10
Gbit/sec.  Infiniband, the newest network fabric, currently is available with
8 Gbit/sec bandwidth in each direction, with 24 Gbit/sec expected in 2005.
Such high bandwidth networks raise the network dispersion curve of
Fig.~\ref{gott_myr} sufficiently to support many forthcoming generations of
processors.

\section{OPTIMIZING PRICE/PERFORMANCE}

Many factors must be taken into consideration when building clusters to
optimize price/performance.  As discussed above, either floating point
performance, memory bandwidth, or communications performance will be the
limiting factor for throughput.  It makes little sense to spend additional
funds for faster nodes or larger clusters if the network fabric limits
performance.  On the other hand, an investment in network hardware with excess
bandwidth can be very cost effective, as the fabric may be reused when nodes
are upgraded or replaced.

\subsection{Node Costs}

At the present time, low cost commodity computers are available with either
one or two processors.  Computers with more than two processors exist, but are
significantly more expensive per processor.  Since network interface cards
represent a significant fraction of the total cost of a tightly coupled
cluster, minimizing the number of interfaces by using dual processor systems
can greatly lower overall costs.  Also, dual CPU systems lower the labor costs
for building and administrating clusters.  On the other hand, single processor
systems as a rule have greater memory bandwidth per processor.

Over the last several years, the most cost effective computers for single node
computations have been single processor machines.  At any given time, the
highest front side bus speeds, 800 MHz currently and soon 1066 MHz, have been
available only on single processor systems.  Furthermore, these systems are sold
in huge volumes as business desktops and home machines, driving prices down.
Their use in clusters, however, was questionable before 2004 because none of
these systems had fast PCI I/O buses.  In 2004, systems with PCI-X and PCI
Express buses entered the market.  In Fermilab's May 2004 purchase of 130
single Pentium 4E systems, the cost per node was approximately \$900 for
systems with server-class motherboards, 2.8 GHz processors, 1 GByte of system
memory, and PCI-X I/O.

Dual processor systems generally cost less than two times the price of
corresponding single processor computers.  Systems based on Intel {\tt ia32}
processors have shared memory buses; the processors in these systems compete
with each other for memory bandwidth, and as a consequence SMP scaling on
lattice codes is poor.  However, these systems tend to have very capable PCI
I/O buses.  The correct approach when high performance I/O is required is to
purchase dual-capable systems populated with only a single processor.

Since mid-2003, dual-processor systems based on AMD's Opteron processor have
been available.  These systems include a memory controller embedded in each
processor as well as distinct local memory buses attached to each CPU.  Access
from an Opteron processor to memory attached to the other CPU is considerably
slower than access to the local memory.  Optimizing lattice codes on these
computers requires modifications to the operating system and user code to take
into account the non-uniform memory architecture.

\subsection{Network Costs}

As a rule, the cost of high performance network fabrics is at least half as
much, and often equal to, the cost of the computing nodes.  Furthermore, distinct
jumps in the cost per node of network fabrics occur as clusters grow in node
count beyond the size of the largest available switch.  Larger clusters
require cascading of switches, with a correspondingly higher cost per switch
port.  Typical costs for non-cascaded switched fabrics based on Myrinet or
Infiniband are approximately \$1000 per node, including the switch, cabling,
and network interface card.  The largest Myrinet switch available at present
has 256 ports.  The largest Infiniband switch has 288 ports.

Lattice QCD clusters with gigabit ethernet mesh fabrics typically have six or
more ethernet ports per node, with each port connected directly to a
neighboring node.  Dual port interfaces are available for approximately \$150
each.  The lower cost of these meshes must be balanced against larger cable
plants, the need for custom communications software, and the sensitivity of
the cluster to node failures.

\section{HISTORICAL TRENDS AND PREDICTIONS} 

\begin{figure}[t]
\includegraphics[width=18.5pc,height=13pc]{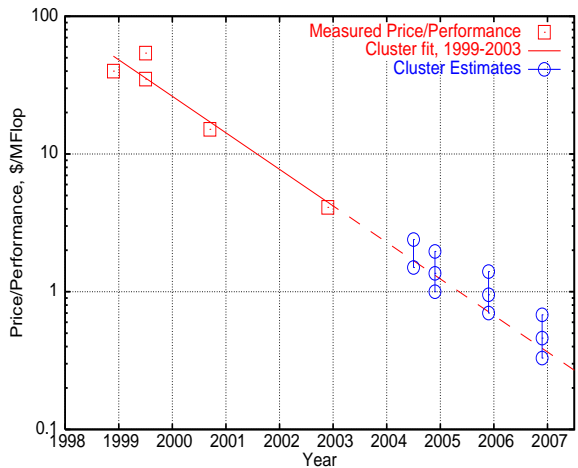}
\caption{Price/performance measurements and predictions for Intel {\tt ia32}
  clusters. Shown are measured (1998 - 2003) and estimated (2004)
  price/performance values for clusters at Sandia National Laboratory, Indiana
  University, and Fermilab, as well as predicted price/performance values for
  clusters to be built in late 2004, 2005, and 2006.  }
\label{predict}
\end{figure}

Figure~\ref{predict} shows the price/performance of MILC improved staggered
code for five clusters built since late 1998, an estimate of price/performance
for the new Fermilab cluster currently being commissioned, and predictions of
price/performance for clusters to be built in the next three years.  The
oldest cluster shown utilized Pentium II processors with 100 MHz memory buses.
The newest existing cluster uses Pentium 4E processors with 800 MHz FSB.  From
the fit to the existing cluster data, the halving time for price/performance
is 1.25 years.

\begin{table*}[htb]
\caption{Price/Performance Predictions. Performance units are GFlop/sec per node.}
\label{tbl_predict}
\begin{tabular}{c c c c c c c}
\hline
Date & Cluster Size & Processor & Performance & Node Cost & Network Cost &
Price/Perf \\
\hline
2004 & 128 & 2.8 GHz P4E & 1.1 & \$900 & \$900 & \$1.64/MFlop \\
Late 2004 & 256 & 3.2 GHz P4E & 1.4 & \$900 & \$1000 & \$1.36/MFlop \\
Late 2005 & 512 & 4.0 GHz P4E & 1.9 & \$900 & \$900 & \$0.95/MFlop \\
Late 2006 & 1024 & 5.0 GHz P4E & 3.0 & \$900 & \$500 & \$0.47/MFlop \\
\hline
\end{tabular}
\end{table*}

Given the historical performance trends, along with vendor roadmaps, we can
attempt predictions of future lattice QCD cluster price/performance.  These
predictions are based upon the following assumptions:

\begin{itemize}
\item Intel {\tt ia32} processors will be available at 4.0 GHz and 1066 MHz FSB in
  2005. 
\item Processors will be available either singly at 5.0 GHz, or in
  dual core equivalence ({\em e.g.}, dual core 4.0 GHz processors) in 2006.
\item Equivalent memory bus speed will exceed 1066 MHz by
  2006 through fully buffered DIMM technology or other advances.
\item The cost of high performance networks such as Infiniband will
  drop as these networks increase in sales volume and the network interfaces
  are embedded on motherboards.
\end{itemize}

The predictions assume that several new technologies are delayed by one year
from their first appearance on current vendor roadmaps.  For example, vendor
roadmaps predict that 1066 MHz memory buses will appear in 2004, dual core
processors in 2005, and fully buffered DIMM technology in 2005.  By year, the
details of the predicted values in Fig.~\ref{predict}, also summarized in
Table~\ref{tbl_predict} are as follows. In mid-2004, the latest Fermilab
cluster used 2.8 GHz P4E systems at \$900/node.  The measured sustained
performance of this cluster varies from approximately 900 to 1100 MFlop/node,
depending upon lattice layout ({\em i.e.}, the number of directions of
communications).  A Myrinet fabric from an older cluster was reused; this
fabric has an estimated replacement cost of \$900 per node.  In late 2004, a
cluster based on 3.4 GHz P4E processors with \hbox{PCI-E} and Infiniband would
sustain 1.4 GFlop/node, based on the faster processors and the improved
communications.  In late 2005, a cluster based on 4.0 GHz processors with 1066
MHz FSB would sustain 1.9 GFlop/node, based upon faster processors and higher
memory bandwidth.  In late 2006, a cluster based on the equivalent of 5.0 GHz
processors with memory bandwidth greater than 1066 MHz FSB would sustain 3.0
GFlop/node.

\section{LIMITS TO PRACTICAL CLUSTER SIZE}

The network fabrics used on clusters limit both achievable performance and
cost effectiveness.  As discussed previously, the largest single high
performance network switches currently available are 288-port Infiniband
switches.  To build a larger cluster based on such a switched network,
cascading of multiple switches is required.  To preserve bisectional bandwidth
through the fabric, switches in a two-layer cascaded fabric have as many
connections to other switches as they do to compute nodes.  Cascading
increases the switch costs of a fabric.


Toroidal gigabit ethernet mesh designs do not have this limitation.  However,
the use of ethernet requires custom communications software to replace the
traditional TCP/IP communications protocol; TCP/IP introduces too much latency
for lattice QCD codes.  In contrast, the communications software which is
supplied with networks such as Myrinet and Infiniband not only is widely used
and robust, but it also requires no modification for lattice QCD.  In terms
of reduced custom software development, significant benefits may be derived
from using popular high performance switched networks, even though the
hardware costs may be greater.

The term ``strong scaling'' refers to the decrease in time to solve a fixed size
problem as additional nodes are employed.  Communications latencies limit
strong scaling.  As node counts increase, the size of the local lattice stored
on each node decreases, and so the size of the messages used to communicate
neighboring hyperplanes also decreases.  Because of the dispersion of
communications bandwidth with message size caused by latency, the decreasing
bandwidth available with shorter messages will eventually limit the
performance as the number of nodes increases.


The reliability of the nodes in a cluster will limit the length of the longest
calculation.  Typical MTBF figures for commodity computers are of order
$10^5$ to $10^6$ hours.  For $10^3$ nodes, an MTBF of $10^5$ hours will
result in an average of one hardware failure every 100 hours.  Operating
system stability may play a role as well, with ``mean time between reboots''
similarly dictating maximum job lengths.  This problem can be addressed by
checkpointing long calculations at regular intervals, so that they may be
restored at an intermediate position after cluster repair.  Note that switched
networks are very tolerant of node failure in that a given sublattice may be
relocated to any available node in the cluster at the start of the next job.
Mesh networks, on the other hand, are generally limited to nearest computer
neighbor communications unless a large latency penalty is incurred.  The loss
of a node within one of the dimensions of a mesh architecture requires
rewiring to route around the failed computer.


\section{CONCLUSIONS}

Since 1999, PC clusters have exhibited steadily improving price/performance
for lattice QCD; the measured price/performance halving time for improved
staggered codes over this time period was 1.25 years.  Performance trends
indicate that balanced designs will be achievable on large scale
clusters in the future.  With the advent of \hbox{PCI-E}, I/O bus designs
will have more than sufficient bandwidth to match the communications
requirements of many future generations of processors.  Networks such as
Infiniband similarly have excess bandwidth today, and vendor roadmaps indicate
performance growth which will pace or exceed processor requirements.  
Improvements in memory designs should provide sufficient memory bandwidth to
balance faster processors.

To date, the largest clusters in the US specifically devoted to lattice QCD
have been no larger than 256 processors and have been based on Myrinet or
gigabit mesh networks.  Based on performance and cost trends, it is clear that
significant clusters will be constructed in the coming years.  A 512 processor
cluster in 2005 should sustain 1.9 GFlop/sec per node on the improved
staggered action at less than \$1/MFlop price/performance.  By 2006, a cluster
with several thousand processors should sustain multiple TFlop/sec per node
for less than \$0.50/MFlop.  Leveraging the results of the wide spread use of
commodity clusters, these facilities will require neither specialized designs
nor operational procedures.

\end{document}